# Electromagnetic wave manipulation using layered systems

**Huanyang Chen[1,2] and C. T. Chan[1*]**

[1]Department of Physics, Hong Kong University of Science and Technology,

Clear Water Bay, Kowloon, Hong Kong, China

[2]Institute of Theoretical Physics, Shanghai Jiao Tong University, Shanghai 200240, China

*e-mail: phchan@ust.hk

## Abstract

We show that the optical properties of an oblique layered system with two kinds of isotropic materials can be described using the concept of transformation media as long as the thickness of the layers is much smaller than the wavelength. Once the connection with transformation media is established, we then show that oblique layered system can serve as a universal element to build a variety of interesting functional optical components such as wave splitters, wave combiners, one-dimensional cloaking devices and reflectionless field rotators.

Anisotropic materials[1] with required permittivity properties can be realized by a layered structure of thin, alternating layers of metal and dielectrics. Such layered structures have applications in many areas, including subwavelength imaging[1], negative refraction[2], optical hyperlenses[3, 4], and invisibility cloaking[5]. These structures are particularly useful for applications at optical frequencies as they can be fabricated precisely[2, 6]. Here, we show that the optical properties of an oblique layered system with two kinds of isotropic materials can be interpreted using the concept of transformation media[7-11]. Once the connection with transformation media is established, it is immediately obvious that oblique layered system can serve as a reflectionless wave shifter[12], which in turn can be used as a universal element to build a variety of interesting functional optical components such as wave splitters, wave combiners, one-dimensional cloaking devices and reflectionless field rotators.

We first establish that the optical properties of an oblique layered system can be interpreted using the concept of transformation media. We begin with an alternating layered system with two kinds of isotropic materials whose permittivities are $\varepsilon_1$ and $\varepsilon_2$. We take the normal direction to be along the $\hat{x}$-direction. The thickness of each layer is identical. In the long wavelength limit, the optical properties of the layered system can be represented by an anisotropic effective medium with an effective permittivity tensor[1] of the form,

$$\vec{\vec{\varepsilon}}_{normal} = \begin{bmatrix} \varepsilon_x & 0 \\ 0 & \varepsilon_y \end{bmatrix} = \begin{bmatrix} \dfrac{2\varepsilon_1 \varepsilon_2}{\varepsilon_1 + \varepsilon_2} & 0 \\ 0 & \dfrac{\varepsilon_1 + \varepsilon_2}{2} \end{bmatrix}. \qquad (1)$$

Here, we consider two-dimensional systems (the $\hat{z}$-direction is homogenous) and transverse electric (TE) polarized (magnetic field in the $\hat{z}$-direction) incident waves.

Let us cut the layered system obliquely[13], as in the region $x_1 < x < x_2$ in Fig. 1a, so that the normal direction of the alternating layers is along the $\hat{u}$-direction at a fixed angle, $\alpha$, from the $\hat{x}$-direction. The effective permittivity tensor of the oblique layered system in $x_1 < x < x_2$ is

$$\ddot{\varepsilon}_{oblique} = \begin{bmatrix} \cos\alpha & \sin\alpha \\ -\sin\alpha & \cos\alpha \end{bmatrix} \begin{bmatrix} \varepsilon_u & 0 \\ 0 & \varepsilon_v \end{bmatrix} \begin{bmatrix} \cos\alpha & -\sin\alpha \\ \sin\alpha & \cos\alpha \end{bmatrix}, \quad (2)$$

where $\varepsilon_u = \varepsilon_x$ and $\varepsilon_v = \varepsilon_y$. The $\hat{v}$-direction is another principal axis perpendicular to the $\hat{u}$-direction.

Here, we show that the optical properties of the configuration depicted in Fig. 1a can be interpreted using the concept of transformation media. Figure 1a shows an obliquely cut layered system that is bounded between two vacuum half spaces. Figure 1b shows a mapping of coordinates, defined by

$$\begin{aligned} &x' = x, \; z' = z \text{ and } y' = y \text{ (for } x < x_1\text{)}, \\ &x' = x, \; z' = z \text{ and } y' = y + t(x - x_1) \text{ (for } x_1 < x < x_2\text{)}, \\ &x' = x, \; z' = z \text{ and } y' = y + t(x_2 - x_1) \text{ (for } x > x_2\text{)}, \end{aligned} \quad (3)$$

where $t$ is a constant.

According to the transformation media concept, the transformation of the coordinates can be mapped to a change of material properties in the original coordinate system, with the permittivity and permeability tensors related by[9]

$$\begin{aligned} \ddot{\varepsilon}' &= |\det(\ddot{\Lambda})|^{-1} \ddot{\Lambda}\ddot{\varepsilon}\ddot{\Lambda}^T, \\ \ddot{\mu}' &= |\det(\ddot{\Lambda})|^{-1} \ddot{\Lambda}\ddot{\mu}\ddot{\Lambda}^T, \end{aligned} \quad (4)$$

where $\ddot{\Lambda} = \dfrac{\partial(x'y'z')}{\partial(x,y,z)}$ is the Jacobian transformation tensor.

In the specific transformation defined by equation (3), we have for $x_1 < x < x_2$

$$\ddot{\varepsilon}' = \begin{bmatrix} 1 & 0 \\ t & 1 \end{bmatrix} \begin{bmatrix} \varepsilon_a & 0 \\ 0 & \varepsilon_b \end{bmatrix} \begin{bmatrix} 1 & t \\ 0 & 1 \end{bmatrix}, \qquad (5)$$

where $\ddot{\varepsilon} = \begin{bmatrix} \varepsilon_a & 0 \\ 0 & \varepsilon_b \end{bmatrix}$ is the permittivity tensor of the base material before the coordinate mapping and $\ddot{\Lambda} = \begin{bmatrix} 1 & 0 \\ t & 1 \end{bmatrix}$ is the transformation tensor for the mapping defined by equation (3). We note that the permeability has the same form as equation (5), but permeability does not affect the propagation of TE waves.

By comparing equations (5) and (2), we find the correspondence between an oblique layered system and the transformation media:

$$t = \dfrac{(\varepsilon_v - \varepsilon_u)\cos\alpha\sin\alpha}{\varepsilon_u \cos^2\alpha + \varepsilon_v \sin^2\alpha}, \quad \varepsilon_a = \varepsilon_u \cos^2\alpha + \varepsilon_v \sin^2\alpha, \quad \varepsilon_b = \dfrac{\varepsilon_v \varepsilon_u}{\varepsilon_u \cos^2\alpha + \varepsilon_v \sin^2\alpha}. \qquad (6)$$

The optical functionality of the oblique layered system can therefore be described, within an effective medium description, using the transformation media concept with the parameters $t$, $\varepsilon_a$ and $\varepsilon_b$.

Now, let $\varepsilon_a = \varepsilon_b = 1$. We therefore obtain

$$\varepsilon_u = (2 + t^2 - t\sqrt{t^2 + 4})/2 \quad , \quad \varepsilon_v = (2 + t^2 + t\sqrt{t^2 + 4})/2 \quad \text{and} \quad \alpha = -\tau/2 \quad \text{with}$$

$\cos\tau = \dfrac{t}{\sqrt{t^2 + 4}}$, $\sin\tau = -\dfrac{2}{\sqrt{t^2 + 4}}$. This means that a layered material with an effective permittivity tensor

$$\ddot{\varepsilon}_{normal} = \frac{1}{2}\begin{pmatrix} (2+t^2-t\sqrt{t^2+4}) & 0 \\ 0 & (2+t^2+t\sqrt{t^2+4}) \end{pmatrix} \quad (7)$$

and its layers aligned obliquely along a direction ($\hat{v}$-direction in Fig. 1a) with slope $k = \tan(\frac{\pi}{2}-\alpha) = \cot(\alpha) = \cot(-\tau/2) = \frac{\sqrt{t^2+4}+t}{2}$ can be used as a reflectionless wave shifter. Light beams propagating through the region $x_1 < x < x_2$ shift by a constant displacement of $t(x_2 - x_1)$ in the $\hat{y}$-direction for light incident from the left. We will call $t$ the "shift parameter".

We note that Rahm *et al.*[12] have extended the transformation media method to designing a complex reflectionless wave shifter. We demonstrate that a very simple oblique layered system can serve as a reflectionless wave shifter, and the functionality of the oblique layers can also be interpreted using the transformation media concept. We show that the wave shifter with oblique layers can be employed as the basic element to form different kinds of reflectionless devices.

To demonstrate the wave-shifting functionality, we let $x_1 = 0$, $x_2 = 1$, and $t = 2$. Consider an TE Gaussian beam incident from the left along the $\hat{x}$-direction with its waist equal to the wavelength. Figure 2a shows the propagation of the incident Gaussian beam in free space. Figure 2b shows that the same Gaussian beam shifts by two wavelengths in the $\hat{y}$-direction after passing through the oblique layered system with no observable reflection at the boundaries. The ratio of the layer thickness to the wavelength is 4.8%, and the layers are $\varepsilon_1 = 0.086$ and $\varepsilon_2 = 11.6$, respectively. In Fig. 2c, we show the propagation of the Gaussian beam through the effective anisotropic medium found using the transformation media method. We see a similar field distribution as shown in

Fig. 2b, reinforcing the notion that we can use the transformation media concept to interpret the functionality of the oblique layered system. The oblique layered system can serve as a reflectionless wave shifter as its impedances are matched at the boundaries and they are intrinsically reflectionless due to the use of continuous transformations within the context of the transformation media. We note that the designs of Rahm *et al.*[12] are based on finite embedded coordinate transformations in which the permittivity tensor becomes position dependent. The oblique layered system is a simpler way to achieve reflectionless beam shifting. More properties of the wave shifter are given in Appendix (1). It is intuitively obvious that this oblique layered system can be utilized as the basic element to design a large variety of reflectionless devices, such as wave splitters and wave combiners, exactly like those suggested by Rahm *et al*[12]. More details can be found in Appendix (2).

Next, we show that the shifter based on the oblique layered system can be used as an element to build more complex devices, such as the one-dimensional cloaking device shown in Fig. 3a. The device has two thin slabs that are back-to-back, each with two shifters that shift light in opposite directions. The four shifters with $t = \pm 3$ ($t = +3$ in regions I and IV, $t = -3$ in regions II and III) act together to open up the cloaking region (region V) defined by ($|y| < 3x + 1.5$ and $|y| < -3x + 1.5$). The unit is the wavelength. These four wave shifters together with a perfect magnetic conductor (PMC) diamond-shaped boundary can function as a cloaking device that reduces scattering in one dimension. Figure 3b shows the passage of a Gaussian beam whose waist is equal to two wavelengths through this one-dimensional cloaking device. Given the equivalence of the layered system and the transformation media, we used the anisotropic medium in the simulation to represent the wave shifters. Figure 3c shows the passage of the same Gaussian beam through a PMC plate with an infinitesimally small thickness and the length equal to one wavelength. By comparing Fig. 3b and Fig. 3c, it is clear that the cloak has the same small scattering cross section as the thin PMC plate has. In fact, we can put any object inside the diamond-shaped cloaking region, and the scattering cross section will be reduced to exactly the same cross section as the thin plate shown in Fig. 3c. Figure 3d shows the scattering pattern of a diamond-shaped object with PMC surfaces (with no shifters), and the scattering is significant. The functionality of the layered system shifter can be interpreted again using the concept of transformation media: the shifters mapped the PMC plate (Fig. 3c) into a diamond-shaped PMC boundary and vice versa.

We note that essentially all realizations of cloaking are so-called "reduced" cloaks[14-16], which are approximations of exact cloaking, as the exact implementation of perfect cloaks requires material properties that are too demanding for fabrication. "Reduced" cloaks are of course not perfect and have some intrinsic scattering cross sections. Here, we show that the layered system configuration combined with the idea of transformation media give us the possibility of achieving a "perfect" rotation cloak[17] that does not require any "reduction", and only two kinds of isotropic materials are required. Although there are many kinds of transformation media designed for novel applications, only a non-magnetic reduced cloak has been designed precisely using the concentric layered system[5]. However, non-magnetic reduced cloaks demand infinitely many kinds of isotropic materials due to the position dependence of the permittivity tensor, making it very complex to realize. We will demonstrate that with a layered system, a real rotation cloak can be designed with just two kinds of isotropic materials.

A rotation cloak is an invisible field rotator that rotates the fields so that the information from inside/outside the cloak will appear as if it comes from a different angle, $\theta_0$. Its functionality is shown in Fig. 4a, which shows that the cloak (marked by white solid lines) rotates the fields at an angle $\theta_0 = \pi/2$, while the cloak itself is invisible as it causes no scattering. Consider the following transformation in cylindrical coordinates: $r' = r$, $\theta' = \theta + g(r)$ and $z' = z$, where $g(b) = 0, g(a) = \theta_0$, and $a$ and $b$ are the inner and outer radii of the rotation cloak. Using the coordinate transformation method[7], we can obtain the permittivity tensor and permeability for a rotation cloak[17, 18] within the cloaking region:

$$\vec{\vec{\varepsilon}}_{cloak} = \begin{bmatrix} 1 & 0 \\ t & 1 \end{bmatrix} \begin{bmatrix} 1 & t \\ 0 & 1 \end{bmatrix} \quad \text{and} \quad \mu_z = 1, \tag{8}$$

where $t = -r \dfrac{dg(r)}{dr}$ and $\vec{\vec{\varepsilon}}_{cloak}$ is expressed in polar coordinates. If we choose $g(r) = \theta_0 \dfrac{\ln(b/r)}{\ln(b/a)}$ so that $t = \dfrac{\theta_0}{\ln(b/a)}$ as in ref.[17], $\vec{\vec{\varepsilon}}_{cloak}$ becomes a position-independent tensor. We see that equation (8) has the same form as equation (5). In order words, a rotation cloak is a wave shifter with polar coordinates.

It is then obvious that we can use the concept of the wave shifter with oblique layers to realize a field rotator using layered systems employing two isotropic materials with dielectric constants, $\varepsilon_1$ and $\varepsilon_2$. In Fig. 1c, we illustrate the geometry of the layered system to produce the rotation cloaking effect. The detailed shape of each layer is described in Appendix (3).

We can rewrite equation (8) in Cartesian coordinates (see Appendix (3)) to determine the principal axis of tensor $\vec{\varepsilon}_{cloak}$ in each point inside the cloak, and we will use $\varepsilon_u$ and $\varepsilon_v$ to denote the corresponding principal values. We note that a layered perfect rotation cloak requires $\varepsilon_1 \cdot \varepsilon_2 = \varepsilon_u \cdot \varepsilon_v = 1$ (see Appendix (3)). The perfect rotation cloaking device can be realized at a particular working frequency if we can find a pair of isotropic materials such that their permittivities are the reciprocal of each other. The above consideration assumes that the background is air ($\varepsilon = 1$). What if the rotation cloak is embedded inside another medium with $\varepsilon_j$? Suppose that we can find two materials with permittivities $\varepsilon_i$ and $\varepsilon_k$ such that $\varepsilon_j \varepsilon_j = \varepsilon_i \varepsilon_k$ and $\varepsilon_i, \varepsilon_j, \varepsilon_k > 1$. Then, the materials with $\varepsilon_i$ and $\varepsilon_k$ can serve as the materials to build the rotation cloak. In this case, we can achieve rotation cloaking at a very broad band because all the permittivities are now larger than one and we can employ dielectrics that have little dispersion and absorption. We note that the broad band functionality is specific to the rotation cloak that does not have singularities within the transformation media context. The working frequencies of the invisibility cloak (either ideal or reduced) cannot achieve broad band operation in this manner because singularity exists in the permittivity tensor component in some positions.

Figure 4a shows the magnetic field distribution near the perfect rotation cloak, with the material parameters obtained using the transformation media concept as described in Ref[17]. The incident TE plane wave is from left to right. The cloaking shell has an inner radius of $a = 0.5$ wavelength and an outer radius of $b = 1$ wavelength. The rotation angle is $\theta_0 = \pi/2$. Figure 4b shows the magnetic field distribution near the "layered rotation cloaking device" described in Fig. 1c with a total of 72 alternating layers. This structure has very little scattering and it has the same rotation effect as shown in Fig. 4a. In addition, we show results with 36 layers in Fig. 4c and 18 layers in Fig. 4d[19]. For more coarse-grained configurations (with a small number of layers and each layer becomes thick compared to wavelength), the effective medium theory will fail. The field rotation effect remains, but the scattering becomes increasingly conspicuous, as shown in Fig. 4d.

In conclusion, we have shown that an oblique layered system can serve as a reflectionless wave shifter that can be deployed as a basic element to manipulate light and whose functionality can be interpreted with the transformation media concept. After identifying the equivalence between the layered system and the transformation media, we can use the transformation media concept to design many wave manipulation devices, such as wave splitters, wave combiners, one-dimensional cloaking devices and invisible field rotators. The results demonstrate the elegance of the transformation media concept and at the same time provide an example in which the concept can be implemented with simple geometries.

Acknowledgement

The authors thank Dr. Y. Lai for useful discussions. This work was supported by a central allocation grant from the Hong Kong Research Grants Council, grant no. HKUST3/06C. The computation resources are supported by the Shun Hing Education and Charity Fund.

**Figure legends**

**Figure 1:** The layered system and the coordinate transformation. a, The geometry of the oblique layered system with two materials. b, The coordinate transformation according to equation (3). c, The geometry of the alternating layered system designed to produce the rotation cloaking effect.

**Figure 2:** Finite element simulation results of the magnetic field distribution near the center of the Gaussian beam. a, The beam is in free space. b, The beam passes through the alternating oblique layered system. c, The beam passes through the corresponding transformation media. The black dotted-dashed lines indicate the boundaries of the shifters.

**Figure 3:** The geometry of the one-dimensional cloak and finite element simulation results of the magnetic field distribution near the center of the Gaussian beam. a, The geometry of the one-dimensional cloak. b, The beam is scattered by the one-dimensional cloak. c, The beam is scattered by the PMC thin plate. d, The beam is scattered by the diamond-shaped PMC cylinder. The red solid lines indicate the PMC's boundaries, the black dotted-dashed lines indicate the boundaries of the shifters.

**Figure 4:** Finite element simulation results of the magnetic field distribution near different kinds of rotation cloaks with TE illumination. a, The object is a perfect rotation cloak, as functionalized by the transformation media concept. b, The object is a layered rotation cloak with 72 layers. c, The object is a layered rotation cloak with 36 layers. d, The object is a layered rotation cloak with 18 layers. The white solid lines outline the interior and exterior boundaries of the cloaks.

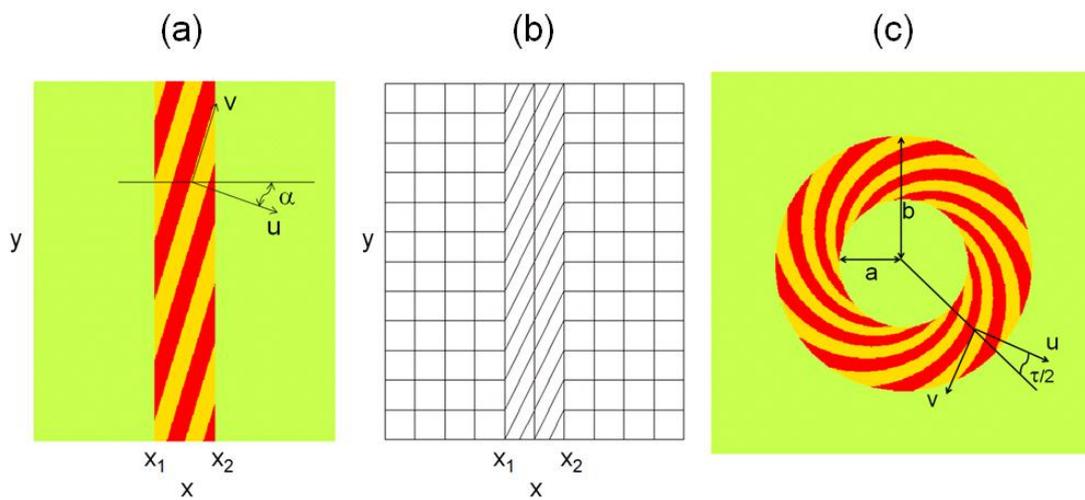

Fig. 1

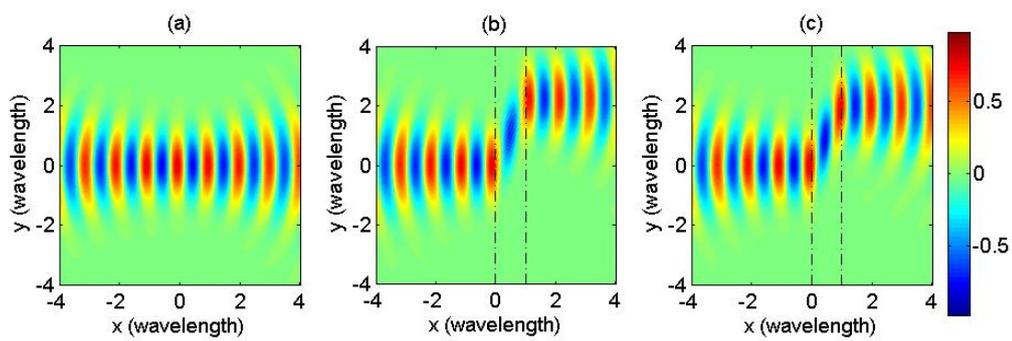

Fig. 2

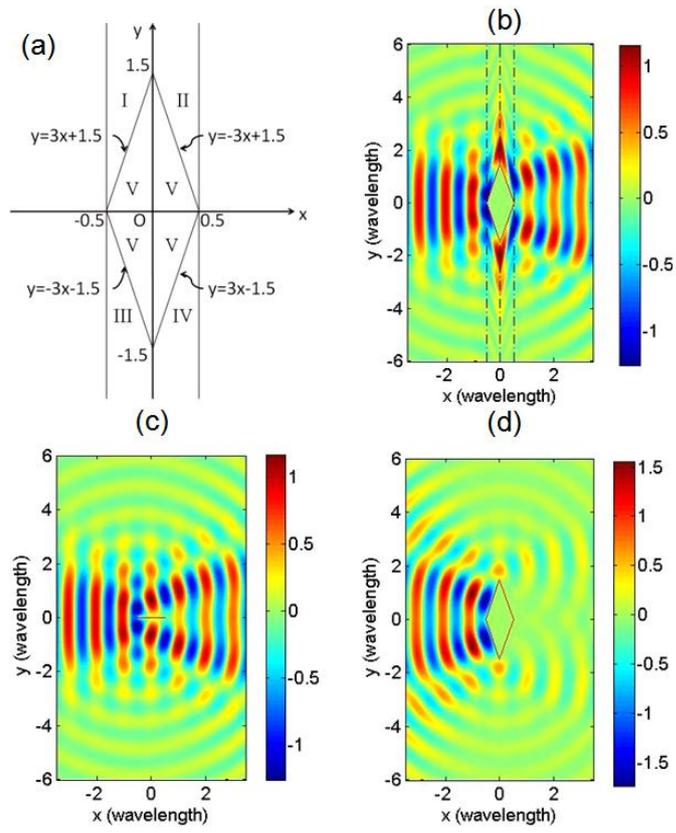

Fig. 3

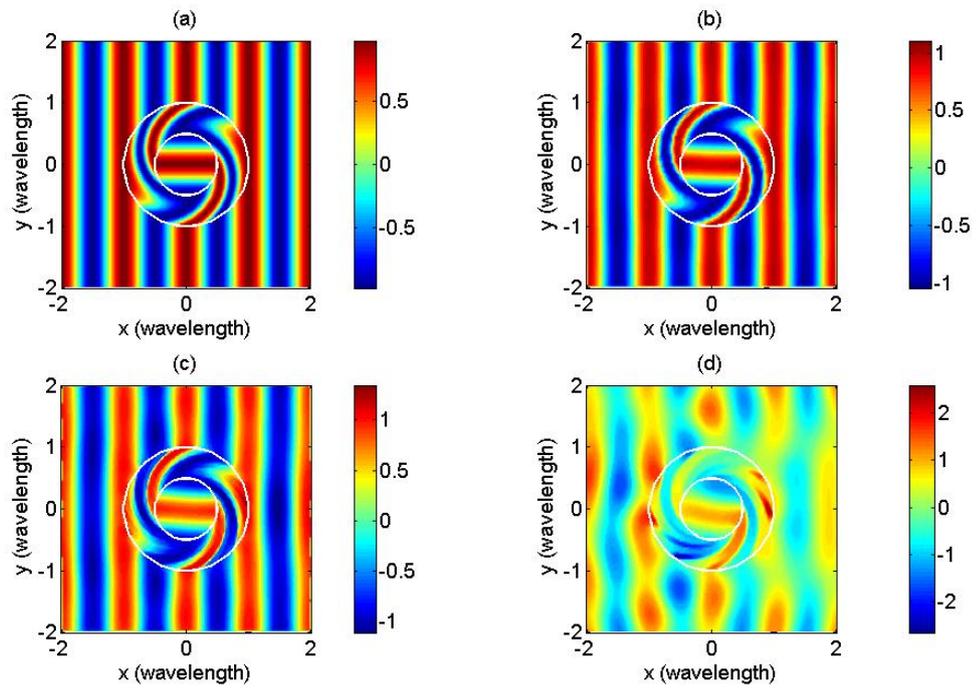

Fig. 4

# Appendix

## (1) The properties of the wave shifter

Let $x_1$ and $x_2$ be the boundaries of the shifter, with $x_1 = -0.5$ wavelength and $x_2 = 0.5$ wavelength. Appendix Figure 1a shows the propagation of an incident Gaussian beam whose waist is equal to two wavelengths from the left to the right along the $\hat{x}$-direction in free space. In Appendix Fig. 1b, we put the wave shifter in the region $x_1 < x < x_2$, with the shift parameter equal to $t = 3$. We see that the Gaussian beam shifts three wavelengths in the $\hat{y}$-direction when it passes through the anisotropic medium. In Appendix Fig. 1c, we put the shifter in the region $x_1 < x < x_2$, but the shift parameter is now $t = -3$, and we see that the Gaussian beam will shift by -3 wavelengths in the $\hat{y}$-direction when it passes through the anisotropic medium.

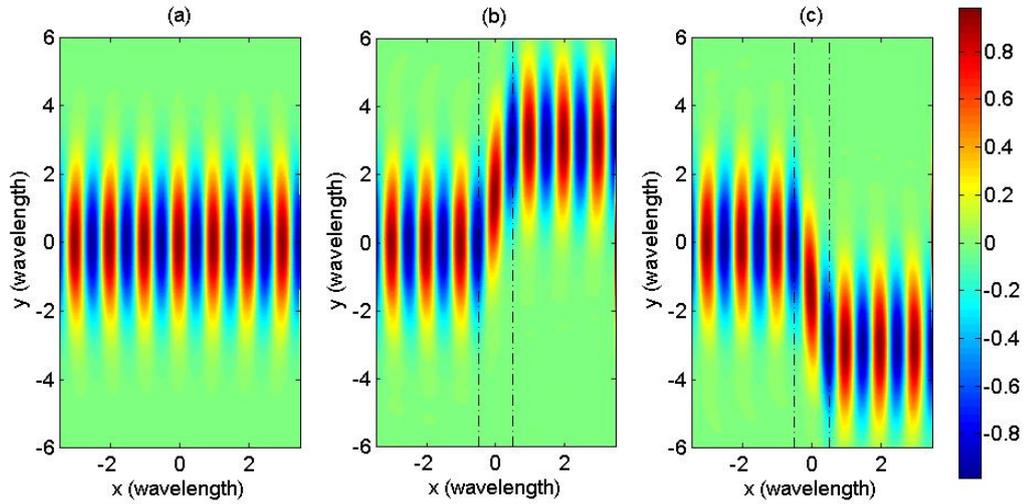

**Appendix Figure 1.** Finite element simulation results of the magnetic field distribution near the center of the normal incident Gaussian beam. a, The beam is in free space. b, The beam passes through the wave shifter with a positive shift parameter. c, The beam passes through the wave shifter with a negative shift parameter. The black dotted-dashed lines indicate the boundaries of the shifters.

Appendix Figure 2a shows the propagation of the incident Gaussian beam from the left to the right but at an angle of 30 degrees from the $\hat{x}$-direction in free space. In Appendix Fig. 2b, we put the shifter in the region $x_1 < x < x_2$, with the shift parameter equal to $t = 3$. We see that the Gaussian beam shifts three wavelengths in the $\hat{y}$-direction when it passes through the anisotropic medium. We note that inside the shifter, the beam propagates as if it were inside a negative refraction index medium. This effect has been discussed in the literature[13]. In Appendix Fig. 2c, we put the shifter in the region $x_1 < x < x_2$, but the shift parameter is now $t = -3$. We see that the Gaussian beam will shift by -3 wavelengths in the $\hat{y}$-direction when it passes through the anisotropic medium.

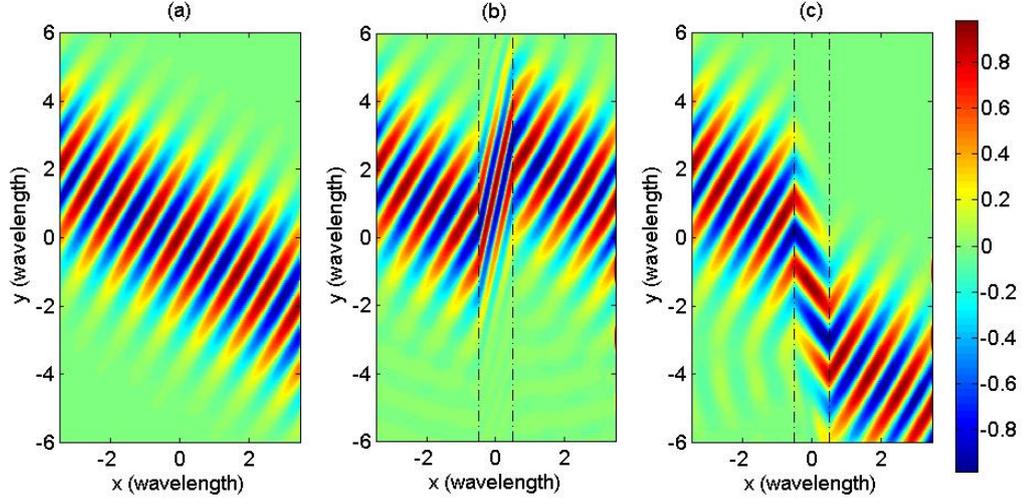

**Appendix Figure 2.** Finite element simulation results of the magnetic field distribution near the center of the oblique incident Gaussian beam. a, The beam is in free space. b, The beam passes through the wave shifter with a positive shift parameter. c, The beam passes through the wave shifter with a negative shift parameter. The black dotted-dashed lines indicate the boundaries of the shifters.

### (2) Wave splitter and wave combiner

The wave shifter can be employed as the basic element to form different kinds of reflectionless devices. If we combine a wave shifter with a positive shift parameter of $t=3$ in the region ($y>0$ and $x_1<x<x_2$) and a wave shifter with a negative shift parameter of $t=-3$ in the region ($y<0$ and $x_1<x<x_2$), we can produce a wave splitter[12] such that one beam can split into two beams after passing through this splitter.

Appendix Figure 3a shows that when the Gaussian beam, whose waist is equal to two wavelengths, has a normal incident with its center at $y=0$, it splits into two beams after passing through the splitter. The splitting beams shift 3 or -3 wavelengths in the $\hat{y}$-direction, and they are interfere with each other to produce the pattern shown. Appendix Figure 3b shows that when the Gaussian beam is obliquely incident at 30 degrees, it will also split into two beams after passing through the splitter, similar to the results shown in Appendix Fig. 3a. If we combine a wave shifter with a negative shift parameter $t=-3$ in the region ($y>0$ and $x_1<x<x_2$) with a wave shifter with a positive shift parameter $t=3$ in the region ($y<0$ and $x_1<x<x_2$), we can produce a wave combiner so that two normally incident Gaussian beams from different positions can combine together into one beam after passing through the combiner. Appendix Figure 3c shows that when these two normal incident Gaussian beams are from $y=3$ and $y=-3$, they will combine together into one beam with the center, $y=0$. We use the anisotropic medium in the simulation in the Appendix (1) and (2) because of the equivalence of the layered system and the transformation media shown above.

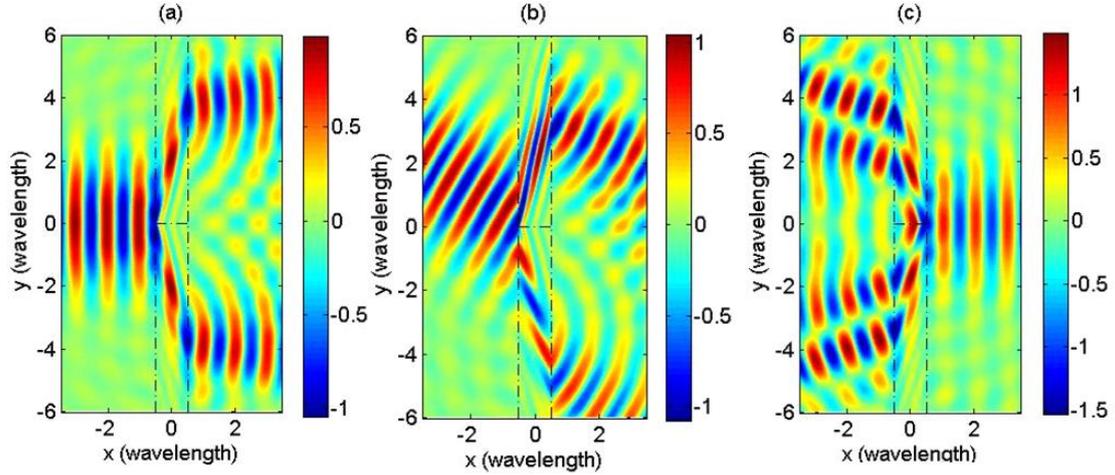

**Appendix Figure 3.** Finite element simulation results of the magnetic field distribution near the center of the incident Gaussian beam(s). a, The beam is normally incident propagating through the wave splitter. b, The beam is obliquely incident propagating through the wave splitter. c, Two beams are normally incident propagating through the wave combiner. The black dotted-dashed lines indicate the boundaries of the shifters.

## (3) The geometry of the alternating layered system to produce rotation cloaking.

Consider the following transformation in cylindrical coordinates: $r'=r$, $\theta'=\theta+g(r)$ and $z'=z$, where $g(b)=0, g(a)=\theta_0$, and a and b are the inner and outer radii of the rotation cloak. Using the coordinate transformation methods[7], we can obtain the permittivity tensor and permeability for a rotation cloak[17] within the cloaking region in Cartesian coordinates:

$$\vec{\vec{\varepsilon}}_{cloak} = \begin{bmatrix} \cos(\tau/2) & -\sin(\tau/2) \\ \sin(\tau/2) & \cos(\tau/2) \end{bmatrix} \begin{bmatrix} \cos\theta & -\sin\theta \\ \sin\theta & \cos\theta \end{bmatrix} \times \begin{bmatrix} \varepsilon_u & 0 \\ 0 & \varepsilon_v \end{bmatrix} \times \begin{bmatrix} \cos\theta & \sin\theta \\ -\sin\theta & \cos\theta \end{bmatrix} \begin{bmatrix} \cos(\tau/2) & \sin(\tau/2) \\ -\sin(\tau/2) & \cos(\tau/2) \end{bmatrix}$$

and

$$\mu_z = 1, \quad (A1)$$

where $\theta$ is the angle in the $(r, \theta, z)$ coordinate, the principal values of $\vec{\vec{\varepsilon}}_{cloak}$ are given by $\varepsilon_u = (2+t^2-t\sqrt{t^2+4})/2$, $\varepsilon_v = (2+t^2+t\sqrt{t^2+4})/2$ ($t = -r\dfrac{dg(r)}{dr}$) and the principal axes ($\hat{u}$-direction and $\hat{v}$-direction) can be obtained by rotating an angle $\tau/2$ from the $\hat{r}$-direction and $\hat{\theta}$-direction with $\cos\tau = \dfrac{t}{\sqrt{t^2+4}}$ and $\sin\tau = \dfrac{2}{\sqrt{t^2+4}}$. If we choose $g(r) = \theta_0 \dfrac{\ln(b/r)}{\ln(b/a)}$ (or $f(r) = \ln(r)$ as in ref.[17]), then $t = \dfrac{\theta_0}{\ln(b/a)}$, and $\varepsilon_u$, $\varepsilon_v$, $\tau$ all become constants. If $\varepsilon_u$, $\varepsilon_v$, $\mu_z$, $\tau$ are chosen as several independent constants, the "reduced" rotation cloak can still rotate fields but it will introduce some scattering so that the cloak becomes visible[20]. This kind of non-perfect "reduced" rotation cloaking has been recently demonstrated theoretically and experimentally[20].

We see that the concentric layered structure with alternating layers of dielectric and metal can be used to mimic the anisotropic properties in the $\hat{r}$-direction and $\hat{\theta}$-direction very precisely if the number of the alternating layers is large enough[3,5]. If we want to use the layered structure concept of the wave shifter to design the perfect rotation cloak, we need to calculate the precise shape of each layer first.

For each point $(x, y)$ in the rotation cloak (see Appendix Fig. 4 a), we imagine a circle (origin at $(0, 0)$) passing through this point. The normal line of this circle indicates the $\hat{r}$-direction while the tangent indicates the $\hat{\theta}$-direction. The slope of the tangential line is $\tan(\alpha) = -x/y$. Since the $\hat{v}$-direction can be obtained by rotating a fixed angle, $\tau/2$, from the $\hat{\theta}$-direction, we can obtain the slope of the line along $\hat{v}$-direction as $\tan(\alpha + \tau/2)$. We image a curve whose tangent is always along the $\hat{v}$-direction of each point, and then we have the differential equation for this curve,

$$\frac{dy}{dx} = \tan(\alpha + \frac{\tau}{2}) = \frac{\tan(\alpha) + \tan(\frac{\tau}{2})}{1 - \tan(\alpha)\tan(\frac{\tau}{2})} = \frac{-\frac{x}{y} + \tan(\frac{\tau}{2})}{1 + \frac{x}{y}\tan(\frac{\tau}{2})} \quad (A2)$$

or

$$\frac{d(r\sin(\theta))}{d(r\cos(\theta))} = \frac{\cos(\theta)r\frac{d\theta}{dr} + \sin(\theta)}{-\sin(\theta)r\frac{d\theta}{dr} + \cos(\theta)} = \frac{-\cos(\theta)\cot(\frac{\tau}{2}) + \sin(\theta)}{\sin(\theta)\cot(\frac{\tau}{2}) + \cos(\theta)} \quad (A3)$$

By comparing the above equations, we obtain a differential equation for this curve,

$$r\frac{d\theta}{dr} = -\cot(\frac{\tau}{2}). \quad (A4)$$

With the starting point in the inner circle, $r = a$, $\theta = \beta$, we have,

$$\theta = \beta + \tan(\frac{\tau}{2})\ln(\frac{a}{r}) \quad (A5)$$

for this curve. Here,

$$\tan(\frac{\tau}{2}) = \frac{1-\cos\tau}{\sin\tau} = \frac{\sqrt{t^2+4}-t}{2} \qquad (A6)$$

We divide the inner circle equally with N (even number) points by setting

$$\beta = 0,\ \frac{2\pi}{N},\ 2\frac{2\pi}{N},\ ...,\ (N-1)\frac{2\pi}{N} \qquad (A7)$$

Staring with these N points, we can produce N curves that will divide the concentric shell ($a<r<b$) into N fan-shaped parts (see Appendix Fig. 4b). We use these alternating "layers" of dielectric and metallic ($\varepsilon_1$ and $\varepsilon_2$) materials the same way as in the aforementioned wave shifter. From the viewpoint of the effective medium theory[1], the device in Appendix Fig. 4b works like a rotation cloak (perfect or non-perfect[20]) with

$$\varepsilon_v = \frac{\varepsilon_1+\varepsilon_2}{2},\ \varepsilon_u = \frac{2\varepsilon_1\varepsilon_2}{\varepsilon_1+\varepsilon_2}. \qquad (A8)$$

To produce a perfect rotation cloak, we can obtain $\varepsilon_1$ and $\varepsilon_2$ inversely,

$$\varepsilon_1 = \varepsilon_v - \sqrt{\varepsilon_v(\varepsilon_v-\varepsilon_u)},\ \varepsilon_2 = \varepsilon_v + \sqrt{\varepsilon_v(\varepsilon_v-\varepsilon_u)}. \qquad (A9)$$

Note that equations (A8) and (A9) are the same as equation (1).

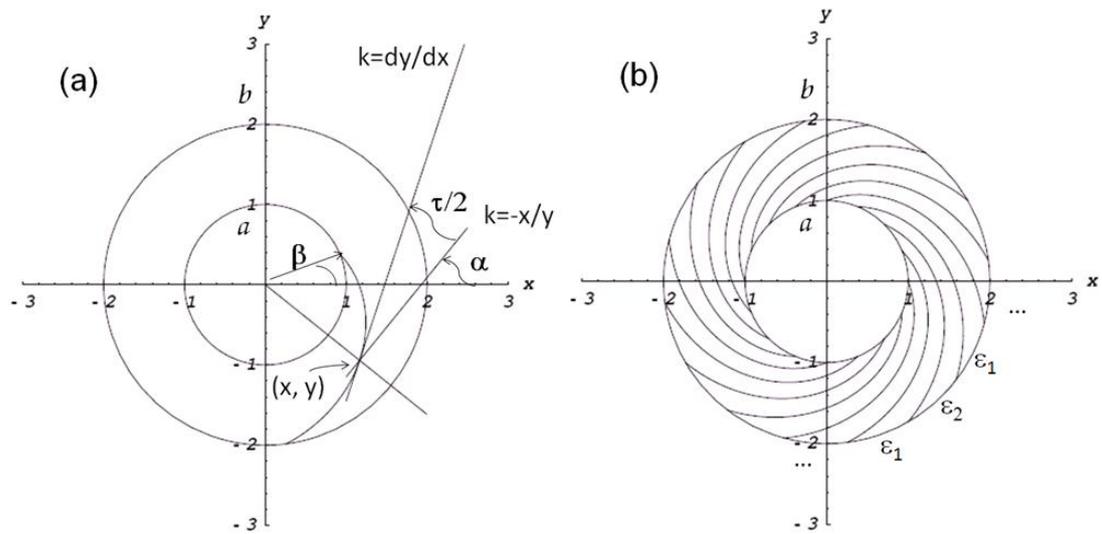

**Appendix Figure 4.** The geometry of the layered rotation cloak. a, The scheme to produce a curve that is mapped from concentric circles. b, Dividing the concentric shell ($a < r < b$) into N layers for the alternating material of dielectric or metals.